\documentclass[journal,twoside,web]{ieeecolor}
\usepackage{generic}
\usepackage{cite}
\usepackage{amsmath,amssymb,amsfonts}
\usepackage{algorithmic}
\usepackage{graphicx}
\usepackage{algorithm,algorithmic}
\usepackage{hyperref}
\hypersetup{hidelinks=true}
\usepackage{textcomp}

\usepackage{booktabs}

\def\BibTeX{{\rm B\kern-.05em{\sc i\kern-.025em b}\kern-.08em
    T\kern-.1667em\lower.7ex\hbox{E}\kern-.125emX}}
\begin{document}
\bstctlcite{IEEEexample:BSTcontrol}

\title{Foundation models for movement data:\\Are they ready for prime-time?}
\author{Alexander Br{\"a}uer, Benjamin Cauchi and %\IEEEmembership{Fellow, IEEE}, 
Nils Strodthoff%, and Third C. Author Jr., \IEEEmembership{Member, IEEE
\thanks{This project received funding by the BMFTR (grant 01KX2419). 
Model training was conducted on infrastructure of the HPC cluster ROSA, located at the University of Oldenburg (Germany). ROSA was funded by the German Research Foundation (DFG) through its Major Research Instrumentation Programme (INST 184/225-1 FUGG) and the Ministry of Science and Culture (MWK) of Lower Saxony.}
\thanks{All authors are with Carl von Ossietzky Universität Oldenburg, Oldenburg, Lower Saxony, 26129, Germany (e-mail: alexander.braeuer@uol.de, benjamin.cauchi@uol.de, nils.strodthoff@uol.de).
}}

\maketitle

\begin{abstract}
Foundation models (FMs) trained on large-scale accelerometer data have been proposed as general-purpose feature extractors for health monitoring, but systematic evidence of their advantages is lacking. We present the first comprehensive evaluation of four open-source accelerometer FMs against supervised baselines covering 19 tasks across the domains of activity recognition including activities of daily living, clinical monitoring, and physiological inference. We find task-dependent performance results:  supervised models remain competitive with FMs on human action recognition (HAR), with no consistent advantage for either, while selected FMs lead on fall and stress detection and are the most robust to sensor-placement variation. As frozen feature extractors, FMs are strongest for demographic inference, whereas sleep staging performance remains near chance level for all models. The internal FM representations show strong similarity across layers, highlighting potential for future FM improvements. Linear and frozen probing reveals that UniMTS provides the strongest representations and is the only FM that surpasses the supervised baselines without finetuning. Concept discovery analysis shows all models capture high-intensity activities clearly but struggle with sedentary, complex or ambiguous activities. We provide scenario-based deployment recommendations. Furthermore, we identify FM-derived activity profile inference—moving beyond fixed category classification—as a promising research direction.
\end{abstract}

\begin{IEEEkeywords}
Activity recognition, Clinical Monitoring, Physiological Inference, Wearable Devices %check with thesaurus %TODO
%Enter key words or phrases in alphabetical order, separated by commas. Using the IEEE Thesaurus can help you find the best standardized keywords to fit your article. Use the thesaurus access request form for free access to the IEEE Thesaurus: \underline{https://www.ieee.org/publications/services/thesaurus-acce}\\
%\underline{ss-page.com.}
\end{IEEEkeywords}

\section{Introduction}
\label{sec:introduction}
\IEEEPARstart{T}{he} rise of foundation models (FMs) has transformed the entire field of machine learning. In this work, we follow the original definition \cite{bommasani2021opportunities} and refer to FMs as large-scale models that have been pretrained on large datasets and can be flexibly adapted to different tasks. The most notable impact of these tasks has been on the domains of computer vision \cite{caron2021emerging} and natural language processing \cite{devlin2019bert}, but the development is now starting to impact also the biomedical domain, where foundation models have been proposed for different medical subdomains from medical imaging \cite{chen2024towards,wu2025towards} over physiological time series \cite{cui2024neuro,al2025benchmarking} to wearable movement data \cite{qiuCustomizableFoundationModels2025a}.

\textbf{Movement data}
In this work, we assess the status of foundation models in the domain of human movement data, i.e., FMs applied to data captured from accelerometers. Acceleration data serve as a valuable source of information for a range of different tasks, most notably human activity recognition (HAR) at various levels of granularity \cite{roggenCollectingComplexActivity2010a,reissIntroducingNewBenchmarked2012,sztylerOnbodyLocalizationWearable2016}, fall detection \cite{sucerquiaSisFallFallMovement2017a,yuLargescaleOpenMotion2021}, sleep/wake or sleep‑stage prediction \cite{logacjovMachineLearningModel2024}, and stress/affect classification \cite{schmidtIntroducingWESADMultimodal2018}. 
As such, movement data from wearables can provide valuable information on the current state of a person and can lead to more informed decisions about the health status of a given person.
Beyond these established tasks, wearable movement data is increasingly adopted in clinical trials \cite{fayadWearableTechnologiesClinical2026}, where accelerometer-derived measures serve both as physiological biomarkers and as passive indicators of patient functioning - capturing physical activity, gait, anxiety and sleep patterns to evaluate treatment outcomes across diverse therapeutic areas \cite{garciaSensorbasedDigitalHealth2026,beltran2025digital}.

\textbf{Research gaps}
Despite the growing number of proposed FMs in the domain, typically trained using self-supervision \cite{logacjovSelfsupervisedLearningAccelerometerbased2024, tangSelfHARImprovingHuman2021}, their systematic evaluation remains a sizable research gap \cite{bianFoundationModelsDefining2026}. First, models are rarely assessed comprehensively, i.e., on a comprehensive set of different tasks and datasets. Secondly, inconsistent evaluation protocols across different publications render model comparisons very difficult. Thirdly, assessments rarely go beyond task performance to analyze computational efficiency, robustness to deployment conditions, or internal representations.

\textbf{Contributions}
In this submission, we aim to close the mentioned research gaps by enabling like-by-like comparisons of FMs on a comprehensive set of relevant tasks, in line with recent benchmarking studies for foundation models in other domains \cite{neidlinger2025benchmarking,al2025benchmarking}. 

More explicitly, we put forward three main contributions namely benchmarking of four FMs, insights into FMs internal representation and recommendations for practitioners:
%the following contributions
%\begin{enumerate}
%\item 
We devise a comprehensive evaluation framework covering 19 tasks from five categories leveraging 10 different datasets. We benchmark four open-weight FMs in three evaluation modes in comparison to strong supervised baselines. 
Our results demonstrate that FMs show no consistent advantage over supervised baselines on standard HAR tasks, while they lead on stress detection and yield the best ready-to-use frozen feature extractors for individual categories (Fig.\ref{fig:model-performance}). %and Table \ref{tab:combined_eval_results_median_ranks}).
We assess model adaptability across seven sensor positions, showing that FMs dominate across most sensor positions (Table \ref{tab:sensor_position}). We compare all models in terms of computational efficiency, revealing order-of-magnitude differences that inform practical deployment decisions (Table \ref{tab:model_efficiency}).

%\item 
We analyze internal representations through layerwise concept-based alignment (CBA) \cite{vielhaben2025beyond} and centered kernel alignment (CKA) \cite{kornblith2019similarity}, revealing a spectrum from homogeneous processing 
%(UniMTS) 
to hierarchical specialization 
%(Oxford SSL/ElderNet), with NormWear in between 
(Fig. \ref{fig:pretrained_layer_cba}). We identify high inter-layer similarities, which hint at suboptimal FM representations. %performance. %Cross-model alignment shows high conceptual similarity across all FMs (Fig. \ref{fig:cross_model_cba}). 
Concept discovery reveals that all models form distinct concepts for dynamic activities but not for sedentary ones (Fig. \ref{fig:UMAP_eldernet}).

%\item 
We conclude with scenario-based deployment recommendations for practitioners grounded in task requirements, computational constraints, and sensor placement variability (Section \ref{recommendation}).
%\end{enumerate}

\section{Materials and Methods}
\subsection{Foundation models and supervised baselines}
In this work, we compare four open-source state-of-the-art FMs. We benchmarked those against four supervised baseline models. Since the benchmark requires access to model weights or at least model access via API, several prominent models in the field, such as SensorLM \cite{zhangSensorLMLearningLanguage2025a}, LSM-2 \cite{xuLSM2LearningIncomplete2025a} and WBM \cite{erturkSensorDataFoundation2025} were unavailable. The four FMs span diverse architectures (transformer, graph-transformer, ResNet) and pretraining strategies (contrastive, MAE reconstruction, multi-task self-supervised) with 5--136~M parameters.

Likewise, the four supervised baselines cover state space models, CNNs, and attention-based architectures with parameters between 117~K--2.2~M.All models are summarized in Table~\ref{tab:models} and described in detail in the Appendix.

\begin{table*}[!ht]
    \centering
    \caption{Model architecture comparison. In this study, we investigate four foundation models (rows 1-4) as well as four baseline models (rows 5-8) trained from scratch. TF = Transformer, LSTM = Long short term memory, ST-GCN = Spatial Temporal Graph Convolutional Network, FS = sampling frequency in Hertz (Hz), IL = pretraining input size in seconds.}%, UKB = UK Biobank}
        \begin{tabular}{llccccc}
            \toprule
            \textbf{Model} & \textbf{Architecture} & \textbf{Pretraining Data} & \textbf{Pretraining Task}  & \textbf{FS} & \textbf{IL}\\
            \midrule
            NormWear & Conv patch+TF & 9 public datasets & MAE reconstruction & 65 & 10\\
            UniMTS & ST-GCN+TF & HumanML3D\cite{guoGeneratingDiverseNatural2022} & Contrastive & 20 & 10\\
            Oxford SSL & ResNet-V2 & UKB\cite{dohertyLargeScalePopulation2017a} & Self-supervised & 30 & 6\\
            ElderNet & Oxford SSL &  UKB\cite{dohertyLargeScalePopulation2017a}, MAP\cite{bennettOverviewFindingsRush2012} & Self-supervised & 30 & 6\\
            \midrule
            TinyHAR & CNN+TF+LSTM &  -- & -- & -- & -- \\
            XResNet1D & 1D ResNet &  -- & -- & -- & --\\
            Inception1D & 1D Inception & -- & -- & -- & --\\
            S4 & State Space Model &  -- & -- & -- & --\\
            \bottomrule
        \end{tabular}
        \label{tab:models}
\end{table*}

\subsection{Datasets and tasks}
%TODO Reason for dataset choice
We leveraged 10 publicly available datasets 
that span five task categories: human activity recognition (PAMAP2, HAR70+, HARTH, RealWorld, USC-HAD, WISDM), fall detection (KFall, SisFall), sleep staging (DualSleep), stress and affection detection (WESAD), and demographics regression tasks (derived from datasets with available metadata). Dataset sizes range from 8 to 38 subjects and 7 to 281 hours of recording, with sampling rates between 20 and 200~Hz. All datasets and their characteristics are summarized in Table~\ref{tab:datasets}, with detailed descriptions in the Appendix.

\begin{table*}[htbp]
    \centering
    \caption{Summary of the datasets utilized in this study. Subjects column shows total (female). Demographics are reported as mean $\pm$ SD where information was available. Subj. = subjects, Cl. = classes, Seg. = segments, Dur. = duration in hours, FS = sampling frequency in Hertz (Hz), demographic tasks indicated with: A = age, W = weight, H = height.}
        \begin{tabular}{llccccccc}
            \toprule
            \textbf{Dataset} & \textbf{Task} & \textbf{Subj. (F)} & \textbf{Cl.} & \textbf{Seg.} & \textbf{Dur.} & \textbf{Age (years)} & \textbf{FS}\\
            \midrule
            PAMAP2 & HAR,A,W & 8 (1) & 10 & 815 & 6.8 & 26.7 $\pm$ 3.2 & 100 \\
            HAR70+ & HAR & 18 & 6 & 1,499 & 12.6 & 70--95 & 50 \\
            HARTH & HAR & 22 & 8 & 4,296 & 35.9 & -- & 50 \\
            RealWorld & HAR & 15 (7) & 8 & 2,020 & 16.9 & 31.9 $\pm$ 12.0 & 50 \\
            USC-HAD & HAR,A,W,H & 14 & 12 & 929 & 7.8 & 30.1 $\pm$ 6.9 & 100 \\
            WISDM & HAR,A,W,H & 51& 18 & 7,986 & 66.7 & 28.3 $\pm$ 10.3 & 20\\
            \midrule
            SisFall & Fall,A & 38 (19) & 34 & 2,590 & 21.7 & 40.2 $\pm$ 21.0 & 200 \\
            KFall & Fall & 32 (0) & 36 & 1,320 & 11.1 & 24.9 $\pm$ 3.7 & 100 \\
            \midrule
            DualSleep & Sleep & 29(17) & 6 & 33,814 & 281.8 & 40.2 $\pm$ 15 & 50 \\
            \midrule
            WESAD & Stress,A,W,H & 15 (3) & 5 & 1,398 & 11.7 & 27.5 $\pm$ 2.4 & 64 \\
            \bottomrule
        \end{tabular}
    \label{tab:datasets}
\end{table*}

\subsection{Procedures}
\textbf{Evaluation modes} We assessed the FM performance in terms of three evaluation modes: linear evaluation, frozen evaluation and finetuning. In linear evaluation, a linear head was applied to the appropriately pooled 
%or CLS-token 
representation of a frozen FM. For frozen evaluation, the linear head was replaced with a learnable query-attention head \cite{bardes2024vjepa}, that uses a single trainable query vector to attend over frozen (unpooled) FM representation tokens. 
Finally, finetuning used the same setup as linear evaluation but 
allowed updating of all FM weights with a uniform learning rate.

\textbf{Training} For comparability, we finetuned all models using a unified pipeline incorporating FMs via appropriate wrapper modules. All models were trained only with triaxial-acceleration data and the wrist sensor position was preferred for all models but NormWear that used chest acceleration channels if available. All FMs used their respective pretraining sampling frequency and settings such as clipping acceleration to $\pm 3 $ gravitational acceleration (g) for the Oxford SSL model or transform input data to acceleration in m/s² for UniMTS otherwise it was transformed to g.

Window-level targets were obtained by a label-aggregation process: for each window we counted the number of tokens belonging to each class and defined the target as the per-class fraction of time the class is present within the window, yielding soft (fractional) multi-label targets. Multi-label classification used 
binary cross-entropy on these fractional targets as optimization target, whereas for regression mean squared error (MSE) was used. Finetuning was done for 100 epochs using AdamW  with a constant learning rate of 0.0003, a weight decay of 0.001, a batch size of 128 and early stopping after 30 epochs of no improvement. Training was validated using macro averaged area under receiver operating characteristic 
(AUROC) %(macro-AUROC)
for classification or mean absolute error (MAE)
%MAE 
for regression. The checkpoint with the best validation score was used for test set evaluation.

\textbf{Evaluation}
To compare and evaluate these models on human activity recognition, the data were split using multi-label stratification based on prediction targets while respecting patient assignments \cite{wagner2020ptb,Sechidis2011} into 10 stratified folds for datasets with 10 or more subjects (8:1:1), 5 stratified folds for datasets with less than 10 subjects (3:1:1). 

Activity classes that could not be equally distributed across the three splits were removed prior to splitting. Because the models operate on fixed input length, each 30 s evaluation window was processed as a set of consecutive, non-overlapping sub-windows matching each models native input length. Each sub-window was scored independently and the resulting per-class predictions were aggregated into a single window-level prediction by max pooling. All reported metrics were then computed on these aggregated window-level predictions.

\textbf{Metrics and statistical significance} 
For multi-label classification, each window's target was the per-class fraction of samples within the window. In order to detect activities within the window, the fractional target was binarized class-wise and a class was treated as positive if it exhibited a non-zero fraction. Consequently AUROC was computed on those binarized ground truth targets per class. Regression tasks (age, weight and height) were scored by MAE between predicted and true values.
To assess statistical significance, model rankings within each task were determined through pairwise bootstrapping on the test set (1000 iterations). Starting from the best model (rank 1), each subsequent model was compared against the current rank leader and shared its rank if the 95\% bootstrap interval of the score difference included zero, otherwise it was ranked worse. This established a ranking table (with potential ties), where differences in rank indicate statistically significant performance differences. 

\textbf{Model efficiency}
Computational efficiency metrics were measured using PyTorch profiler (PyTorch 2.9) on a single NVIDIA L40S graphical process unit (GPU). Giga floating point operations per second (gigaFLOPS) were obtained using profilers 'with\_flops' option. Throughput (samples/s), latency (ms/sample), and peak GPU memory was sampled using 100 inference iterations with a batch size of 64. GPU was synchronized for accurate timing beforehand.

\textbf{Representational similarity and concept discovery} In order to compare the internal representations of the four foundation models, we applied the NLMCD-ALIGN framework \cite{vielhaben2025beyond} for concept discovery and layerwise concept alignment. For each model, backbone features were extracted from the penultimate layer for all RealWorld windows (N=119,019). Non-Linear Multivariate Concept Discovery (NLMCD) was then applied independently per model: Uniform Manifold Approximation and Projection (UMAP) \cite{mcinnesUMAPUniformManifold2020} projected the features into a lower dimensional embedding space, followed by hierarchical density-based spatial clustering of applications with noise (HDBSCAN) density-based clustering \cite{mcinnes2017hdbscan} for cluster discovery. Concept proximity scores were derived via soft membership probabilities. 
We computed pairwise CBA, whereby corresponding concepts were identified to quantify representational agreement between models.
The resulting symmetric CBA matrix captures which models develop similar internal concept structures. % despite differing architectures and pretraining objectives. 
We complement CBA with intra-model layerwise analyses using both CBA and CKA \cite{kornblith2019similarity} to assess how representations evolve across layers.
The extracted features were embedded in a two-dimensional space for visualization of the discovered concepts. Subsequently, the embeddings were then displayed as scatter plot.

\section{Results}
\subsection{Performance evaluation}
\subsubsection{Finetuning performance}
Here, in this section we focus on the finetuning results, whereas frozen/linear evaluation along with insights from hidden representations is discussed in Section~\ref{sec:linearfrozen}.
We summarize the model performance in radar plots covering the three evaluation modes (Fig.~\ref{fig:model-performance}). We stress that differences in rank correspond to statistically significantly different performance. We further aggregated ranks according to task categories to provide a concise summary (Table~\ref{tab:combined_eval_results_median_ranks}). 

In the following, the finetuning performance results are described for the different tasks groups namely HAR, falls, sleep, stress and demographics:
%This section focuses on finetuning results. Frozen/linear evaluation along with insights from hidden representations is discussed in Section~\ref{sec:linearfrozen}.

\begin{figure*}[!ht]
    \centering
    \includegraphics[width=\textwidth]{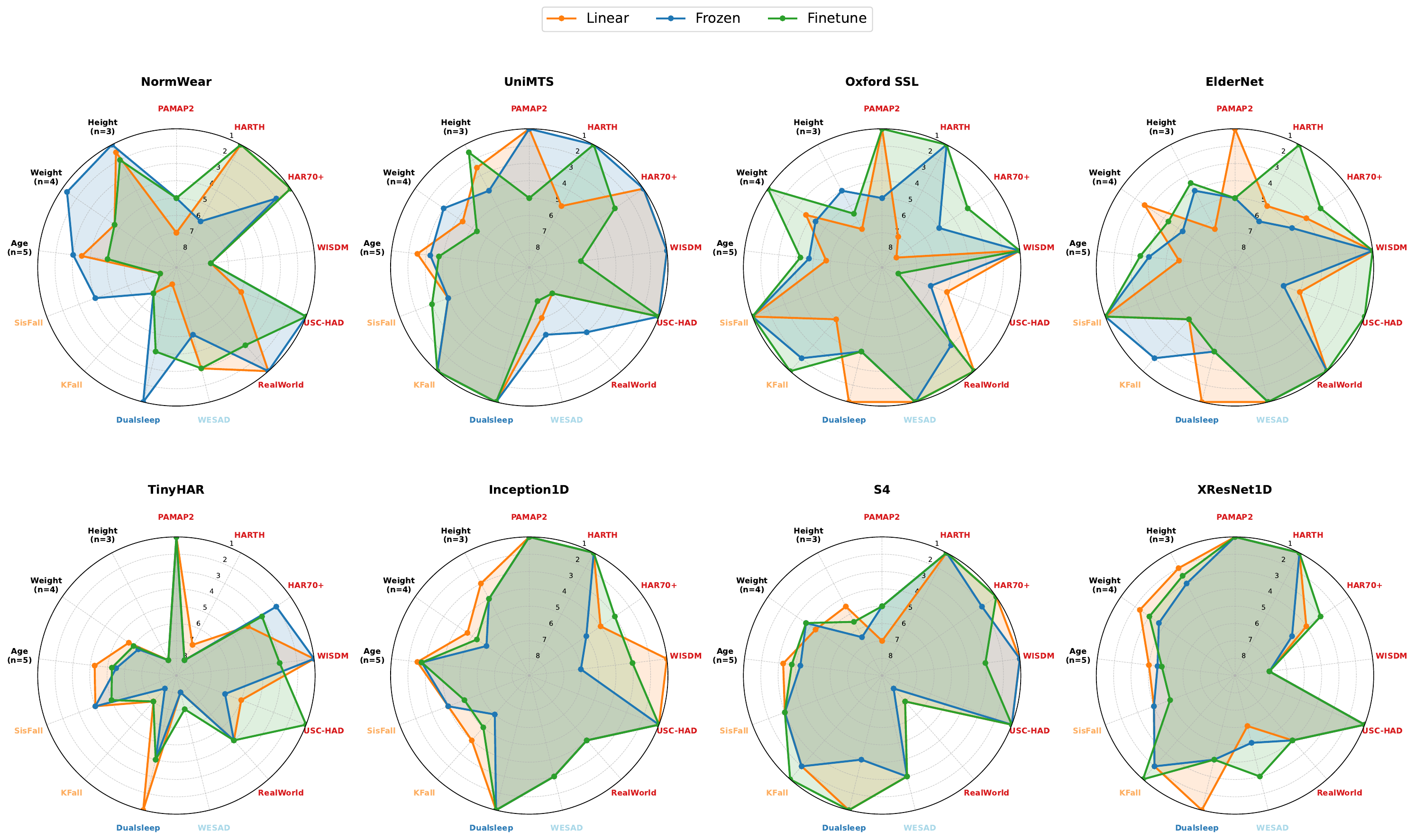}
    \caption{Radar plots summarizing performance ranks for four FMs and four supervised baselines. For similar demographic tasks datasets were combined and the plots show the mean rank. Different ranks indicate statistically significant performance differences, equal ranks are assigned in case of statistically indistinguishable performances. Supervised models were trained only once from scratch but ranked in comparison to FMs for the three evaluation modes.}
    \label{fig:model-performance}
\end{figure*}

%\begin{enumerate}
%\item 
\textbf{HAR} FMs dominate HAR in finetuning, led by Oxford SSL and ElderNet which both rank first on four datasets. Whereas the best performing baseline models (Inception1D, S4, XResNet) perform similarly to NormWear with first rank or statistically tied to first rank on three out of six datasets. UniMTS ranks last across the HAR task, making it the least optimal model for downstream tasks if labels are available and the sensor position cannot be chosen. The performance difference between Oxford SSL and ElderNet indicates that the additional pretraining of Oxford SSL model improves the performance on USC-HAD but diminishes the performance on PAMAP2. The overall performance across all HAR tasks ranges from 0.84--0.987.

%\item 
\textbf{Falls} Oxford SSL is the strongest model for fall detection, ranking first on both datasets, followed by UniMTS and the S4 baseline. In contrast, NormWear ranks last on both datasets. Interestingly, ElderNet's performance varies drastically between both datasets, although KFall was recorded in a similar fashion to SisFall. The overall performance across the fall task ranges from 0.889--0.990.

%\item 
\textbf{Sleep} The overall performance on sleep classification remains poor. Model ranks vary between first and statistically different second place. AUROC ranges between 0.524 on ElderNet to 0.621 on Inception1D. For sleep classification baseline models perform similar or better than the FMs with UniMTS as best-performing FM.%ElderNet and Oxford SSL clearly outperform all other models on sleep staging, while S4 ranks last.

%\item 
\textbf{Stress} Oxford SSL and ElderNet lead on stress detection, outperforming all other models. However, absolute performance remains limited (best-performing model: AUROC 0.66) across all models, suggesting that data from accelerometers alone may be insufficient for reliable stress detection.

%\item 
\textbf{Demographics}
On all demographic prediction tasks, the Inception1D baseline model outperforms all FMs and the other baselines. No model consistently outperforms all other across the demographic task categories: On weight prediction, Oxford SSL outperforms all others. On the other hand, for age and height prediction, the best models and statistically tied ranks are shared between foundation and baseline models. Except for weight prediction there is no clear best performing model within each sub-task-category.

%\end{enumerate}

\textbf{Comparison} No finetuned FM outperforms all other models across all task categories. Aggregated over all tasks, Oxford SSL and ElderNet achieve the best ranks - indicating advantages over the other FMs if sufficient labeled data is available. The FMs mostly lead on activity related tasks (e.g. HAR and falls) but on sleep detection or demographic related tasks the supervised baselines are as competitive or superior, with Inception1D outperforming all FMs on demographics and S4 and Inception1D matching or exceeding every FM on sleep. For sleep and stress the absolute performance remains low for all models, indicating limitation through modality rather than through model choice. Notably, the two strongest finetuned FMs, Oxford SSL and ElderNet, are also the two pretrained on triaxial acceleration data.

These comparisons should be read under the limitation that our benchmark uses triaxial acceleration as a common input modality to enable like-for-like evaluation. NormWear 
as well as UniMTS were pretrained on richer input modalities (multimodal physiological signals and gyroscope data, respectively), thus they are evaluated below their modality-native potential. In contrast to this, Oxford SSL and ElderNet were both pretrained on wrist-worn triaxial acceleration, hence they are less affected.

\subsubsection{Sensor position} 
The ability to adapt to different sensor positions is a crucial property for movement FMs. We study this property based on the RealWorld dataset by finetuning FMs and baseline models on seven distinct sensor positions. The results are compiled in Table \ref{tab:sensor_position}. The sensor placements yielding the highest mean scores are chest and waist, while wrist and head yield the lowest.
Overall FMs lead compared to baseline performance (mean AUROC of 0.949 vs 0.936): On six out of seven positions ElderNet is ranked best or statistically tied to the best model. Only on the upper-arm location ElderNet is outperformed by NormWear and UniMTS. The weakest FM across all sensor position is NormWear ranking first or statistically tied to first rank only on two position (upper-arm and waist). Noticeably, the performance on the waist position is statistically tied across all FMs, XResNet1D and S4 (ranked first) baseline models. 

The performance of the baseline models remains limited: only on three positions (chest, upper-arm and waist) S4 ranked first or is tied to the first ranked model, while other baseline models such as TinyHAR and Inception1D never rank first. Notably, positions with higher overall AUROC show a smaller inter-model differences (e.g. waist), whereas lower-performing placements exhibit larger differences (e.g. wrist). 

To summarize, FMs outperform the supervised baselines across most sensor positions, making them a better choice than supervised baselines for cases in which the sensor position cannot be controlled.

\begin{table*}[!ht]
    \centering
    \caption{Statistical rankings across seven sensor locations for finetuning on selected sensor positions in the RealWorld dataset. The best-performing model for each sensor position is underlined. Models that do not perform statistically significantly worse are set in bold face.}
    \begin{tabular}{ccccc|cccc}
        \toprule
        \textbf{Position} & \textbf{NormWear} & \textbf{UniMTS} & \textbf{Oxford SSL} & \textbf{ElderNet} & \textbf{TinyHAR} & \textbf{Inception1D} & \textbf{S4} & \textbf{XResNet1D}\\
        \midrule
        Chest & 0.940 & \textbf{0.985} & \textbf{0.979} & \textbf{0.984} & 0.965 & 0.974 & \textbf{\underline{0.987}} & 0.967 \\
        Wrist & 0.892 & 0.861 & \textbf{0.959} & \textbf{\underline{0.963}} & 0.886 & 0.889 & 0.848 & 0.898 \\
        Head & 0.892 & \textbf{0.949} & 0.926 & \textbf{\underline{0.962}} & 0.900 & 0.905 & 0.926 & 0.919 \\
        Shin & 0.948 & 0.931 & \textbf{0.982} & \textbf{\underline{0.983}} & 0.957 & 0.945 & 0.970 & 0.949 \\
        Thigh & 0.928 & 0.901 & 0.949 & \textbf{\underline{0.970}} & 0.926 & 0.915 & 0.925 & 0.933 \\
        Upperarm & \textbf{\underline{0.965}} & \textbf{0.950} & 0.942 & 0.946 & 0.944 & 0.906 & \textbf{0.947} & \textbf{0.957} \\
        Waist & \textbf{0.976} & \textbf{0.978} & \textbf{0.970} & \textbf{0.974} & 0.964 & 0.949 & \textbf{\underline{0.979}} & \textbf{0.973} \\
        \bottomrule
    \end{tabular}
    \label{tab:sensor_position}
\end{table*}

\subsubsection{Model Efficiency}
The models were benchmarked in terms of computational complexity (gigaFLOPS), GPU memory usage and inference efficiency (throughput and latency). Oxford SSL and ElderNet are the most efficient with similar profiles: 92~MB GPU memory, 17,500 samples/s throughput and less than 0.1~ms latency. UniMTS ranks third, while NormWear is by far the least efficient, roughly $180\times$ slower and $45\times$ more memory-intensive than Oxford SSL and ElderNet. 
Notably, UniMTS (5~M parameters) is less efficient than Oxford SSL (10~M), demonstrating that parameter count alone does not determine inference cost. 

Among supervised baselines, the CNN-based Inception1D and XResNet1D achieve the highest throughput overall, while S4 and TinyHAR are comparatively slower.
For resource-constrained or real-time deployment, Oxford SSL and ElderNet are the most practical FMs, with efficiency on par with lightweight supervised baselines. NormWear's higher computational demands limit it to offline processing.

\begin{table*}[!ht]
    \centering
    \caption{Comparison of computational cost, memory usage and inference efficiency for supervised baseline models and FMs. The computed metrics are GFLOP forward (F) and backward (B) propagation, peak GPU memory was measured in Megabytes and throughput (Thr) was measured in samples/s whereas latency (Lat) is measured in ms/sample. The respective original pretraining sampling frequency (FS) was used for each of the FMs. In contrast, native dataset FS was used for baseline models. }
    \begin{tabular}{cccccc}
        \toprule
        \textbf{Model} & \textbf{FS [Hz]} & \textbf{Parameters [M]} & \textbf{↓ GFLOP (F /B)} & \textbf{↓ GPU Memory [MB]} & \textbf{↑ Thr / ↓ Lat} \\
        \midrule
        NormWear & 65 & 136.131 & 286.1 / 0.00003688 & 4259.965 & 110.294 / 9.067\\
        UniMTS & 20 & 5.185 & 8.776 / 0.5799 & 779.680 & 2084.980 / 0.480\\
        Oxford SSL & 30 & 10.986 & 0.001187 / 0.002114 & 92.084 & 17696.684 / 0.057\\
        ElderNet & 30 & 11.391 & 0.0006589 / 0.001057 & 93.703 & 17506.913 / 0.057\\
        \midrule
        TinyHAR & 100 & 0.117 & 0.1122 / 0.03783 & 134.782 & 1029.459 / 0.971\\
        Inception1D & 100 & 0.474 & 0.0001577 / 0.0004690 & 181.221 & 15061.698 / 0.066\\
        S4 & 100 & 2.245 & 2.533 / 5.050 & 944.135 & 1911.813 / 0.523\\
        XResNet1D & 100 & 0.890 & 0.0002732 / 0.00008859 & 98.580 & 11694.199 / 0.086\\
        \bottomrule
    \end{tabular}
    \label{tab:model_efficiency}
\end{table*}

\subsection{Insights from hidden representations}
In this section, we go beyond benchmarking FM as black box models by analyzing their hidden representations for more detailed insights.

\subsubsection{Linear/frozen evaluation}
\label{sec:linearfrozen}
%\colnst{PUNCHLINE: probing to assess which concepts were understood by FMs}
Linear/Frozen evaluation can be understood as probing of the hidden representations to assess how  well FMs capture specific concepts implicitly defined via the prediction tasks introduced above. The results are summarized in Fig. \ref{fig:model-performance} with full results in Tables \ref{tab:lin_frozen} and \ref{tab:non_lin_frozen} in the Appendix. In the following, the results of linear and frozen evaluation are described for the task groups namely HAR, falls, sleep, stress and demographics:

UniMTS' representations show exceptionally strong performance in four and five out of six HAR related tasks for linear and frozen evaluation, respectively. Under linear probing the FMs representations are not superior to baseline models trained from scratch: S4 and Inception1D both rank first or statistically tied to first rank on four HAR tasks. In particular on HARTH and PAMAP2 baseline models (S4 and XResNet1D) beat FMs linear representations, and where an FM leads (UniMTS on USC-HAD) the difference to the best baseline is small. In light of UniMTS performance under finetune evaluation where it was ranked last, its frozen and linear representations rank far above its own finetuned model performance, indicating sensitivity to downstream finetuning and representation collapse might be mitigated through a two-stage finetuning approach \cite{Mehari:2021Self}.

On falls, leadership flips between the two datasets: UniMTS gives the best representations on KFall in both modes (0.993 linear, 0.994 frozen) -- consistent with its graph-based architecture, which models spatial relations across body joints, and its motion-capture pretraining -- whereas Oxford SSL and ElderNet lead on SisFall (0.980--0.986). Across the FMs representations, NormWear's are the least suited for fall related tasks.

On sleep, absolute performance remains poor across all evaluation modes. It improves slightly under frozen evaluation, though it still only matches baseline performance. NormWear benefits the most under frozen evaluation, it improves from tied-last under linear probing (0.569) to near-best under the frozen head (0.623).

On stress, Oxford SSL and ElderNet clearly lead in both modes (0.66) while all other models stay near chance and UniMTS even below (0.47). 

On demographics no model dominates: under linear probing the best per sub-task is shared between baselines and FMs, but under the frozen head NormWear is consistently strongest (weight on USC-HAD 10.6~kg and WESAD 2.4~kg, height on USC-HAD 8.7~cm). Strikingly, Oxford SSL and ElderNet collapse on weight and height regression under linear probing (MAE up to 35~kg and 98~cm) and recover only once the non-linear query head is added.

For Oxford SSL and ElderNet their representations seem to converge resulting in similar performance across all tested tasks in both evaluation modes, indicating weak difference between learned representations after additional training of the Oxford SSL model, making them similarly good feature extractors for tested task categories.

For linear probing representations from supervised baselines (XResNet1D) perform across all tasks best. With a non-linear head under frozen evaluation, FMs representation capture task related concepts better than the baseline models.

To summarize, no model gives the best frozen representations across all categories, consequently the extractor should be chosen per task. For HAR, UniMTS is the strongest frozen FM but only matches the supervised baselines under linear evaluation and slightly surpasses them using a non-linear head. For falls, NormWear is least suited and all other FMs perform similarly. For demographics, NormWear with a non-linear head is the most reliable, whereas Oxford SSL and ElderNet fail under linear probing. On sleep and stress absolute performance remains poor for all models, with Oxford SSL and ElderNet the only clear leaders (stress). Under a linear probe, the supervised baselines perform best, whereas the non-linear head lets the FMs capture task-related concepts better than the baselines. Consequently, FM representations should be paired with a non-linear read-out head.

\subsubsection{Representational similarities}
In the previous section we analyzed the representations of the last layer via probing. However, this does not give us insights into the way the information is processed throughout the architecture. We assess the latter through CKA/CBA alignment analyses.

\begin{figure*}
    \centering
    \includegraphics[width=1\linewidth]{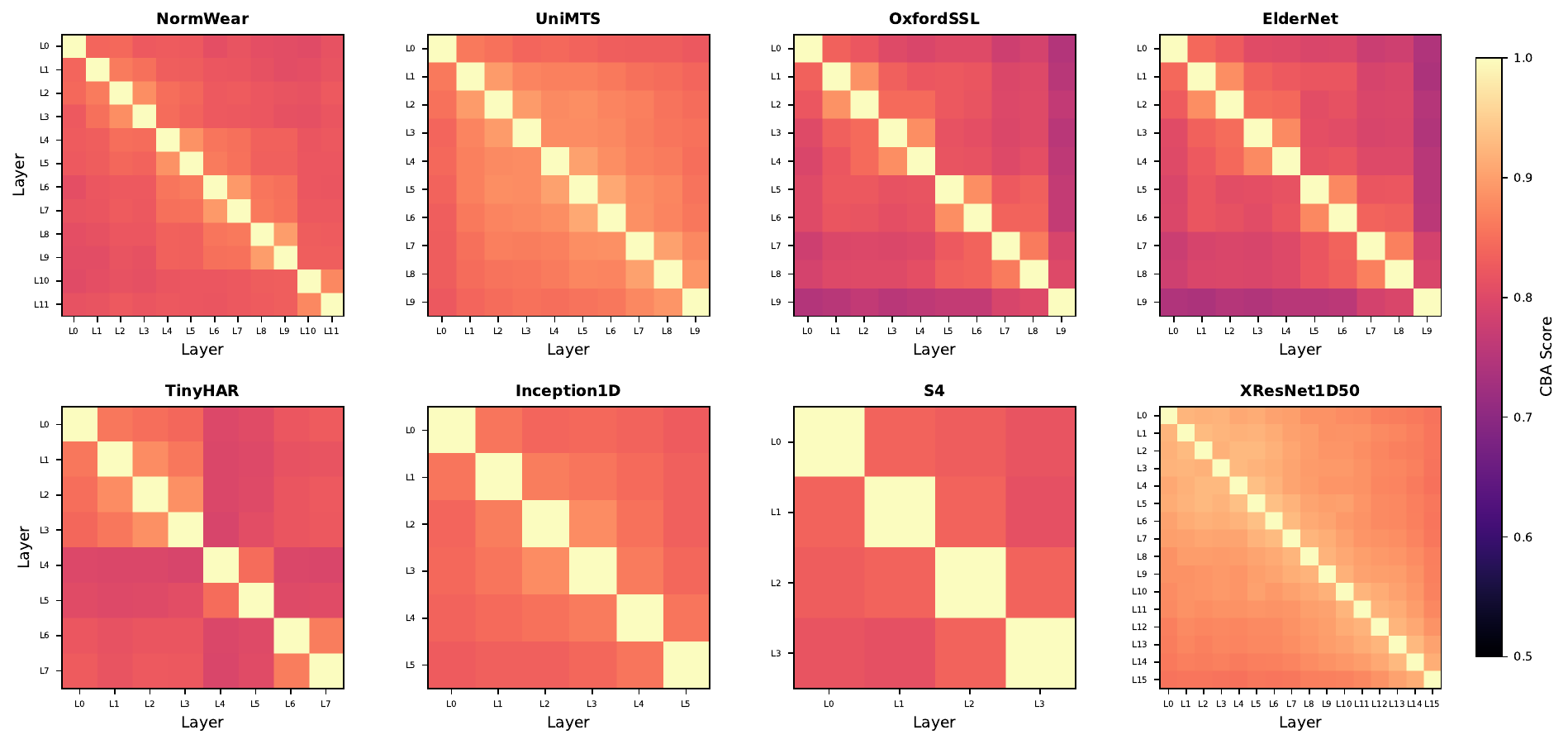}
    \caption{Intra-model layerwise CBA on RealWorld for the finetuned FMs and baseline models. Each heatmap shows pairwise CBA scores between layers of the same model, lighter colors indicate higher concept similarities.}
    \label{fig:finetuned_layer_cba}
    %\Description{Heatmaps showing for each model CBA scores layerwise. CBA scores ranges from 0.73 to 1.0.}
\end{figure*}

Fig. \ref{fig:finetuned_layer_cba} depicts intra-model layerwise CBA scores of the finetuned models. UniMTS shows largely uniform scores across layer pairs, with only minor differentiation emerging between its input layer (L0) and later layers. NormWear displays an intermediate pattern with moderate differentiation, most notably between the first layers (L0--L4) and its final layers L10--L11. %in its final layers (L10--L11). 
In contrast, Oxford SSL and ElderNet exhibit pronounced block-diagonal patterns, where early and late layers encode substantially different concepts—most strongly in ElderNet, with CBA scores ranging between 0.74 to 0.88 for distant layer pairs.%dropping below 0.6 for distant layer pairs. 

\begin{figure*}
    \centering
    \includegraphics[width=1\linewidth]{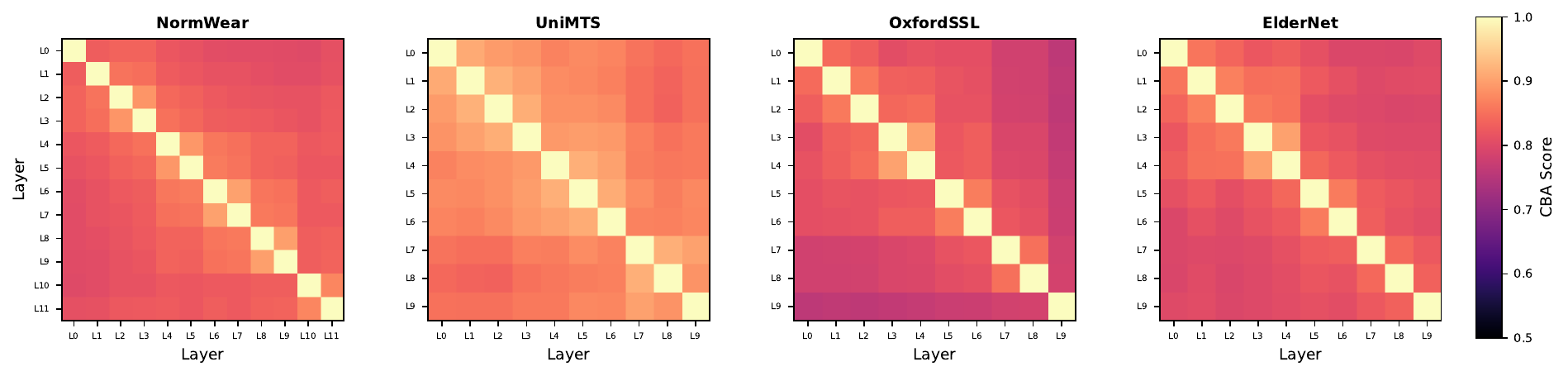}
    \caption{Intra-model layerwise CBA on RealWorld for the FMs using their pretrained checkpoints. Each heatmap shows pairwise CBA scores between layers of the same model, lighter colors indicate higher concept similarities.}
    \label{fig:pretrained_layer_cba}
    %\Description{Heatmaps showing for each FM CBA scores layerwise. CBA scores ranges from 0.6 to 1.0.}
\end{figure*}

Fig. \ref{fig:pretrained_layer_cba} shows intra-model layerwise CBA scores of the pretrained models, i.e., before a potential finetuning step: UniMTS shows the most shallow concept differentiation across layers, whereas Oxford SSL shows the largest and most pronounced differentiation across all layers (CBA scores range from 0.7593--0.9017). Interestingly, compared to Oxford SSL, ElderNet seems to have developed more limited concepts across layers although further pretrained, indicating the pretraining might have successfully resulted in a model specialized for a subpopulation (elderly). The linear CKA results support these findings (Appendix Fig. \ref{fig:cka_fms}). 

For all four FMs, the representational similarity across layers is unexpectedly high, indicating weak layer differentiation due to suboptimal architecture, training design choices or dataset properties. This is particularly striking for UniMTS, where different layers are hardly distinguishable based on representational similarity.

\subsubsection{Insights from concept discovery}
As final analysis, we revisit the final layer hidden representations and look into the alignment of representational substructures with activity labels. As substructures, we consider the concepts discovered during the CBA analysis, which involves dimensionality reduction via UMAP and hierarchical clustering using HDBSCAN. On the one hand, since we are considering FM representations after pretraining, the FM was never exposed to activity labels and therefore the discovered concepts do not have to align with man-made activity labels. On the other hand, potentially fine-grained FM concepts discovered in a data-driven manner could lead to a more detailed categorization of movement states and beyond.

\begin{figure*}[t]
    \centering
    \includegraphics[width=1\linewidth]{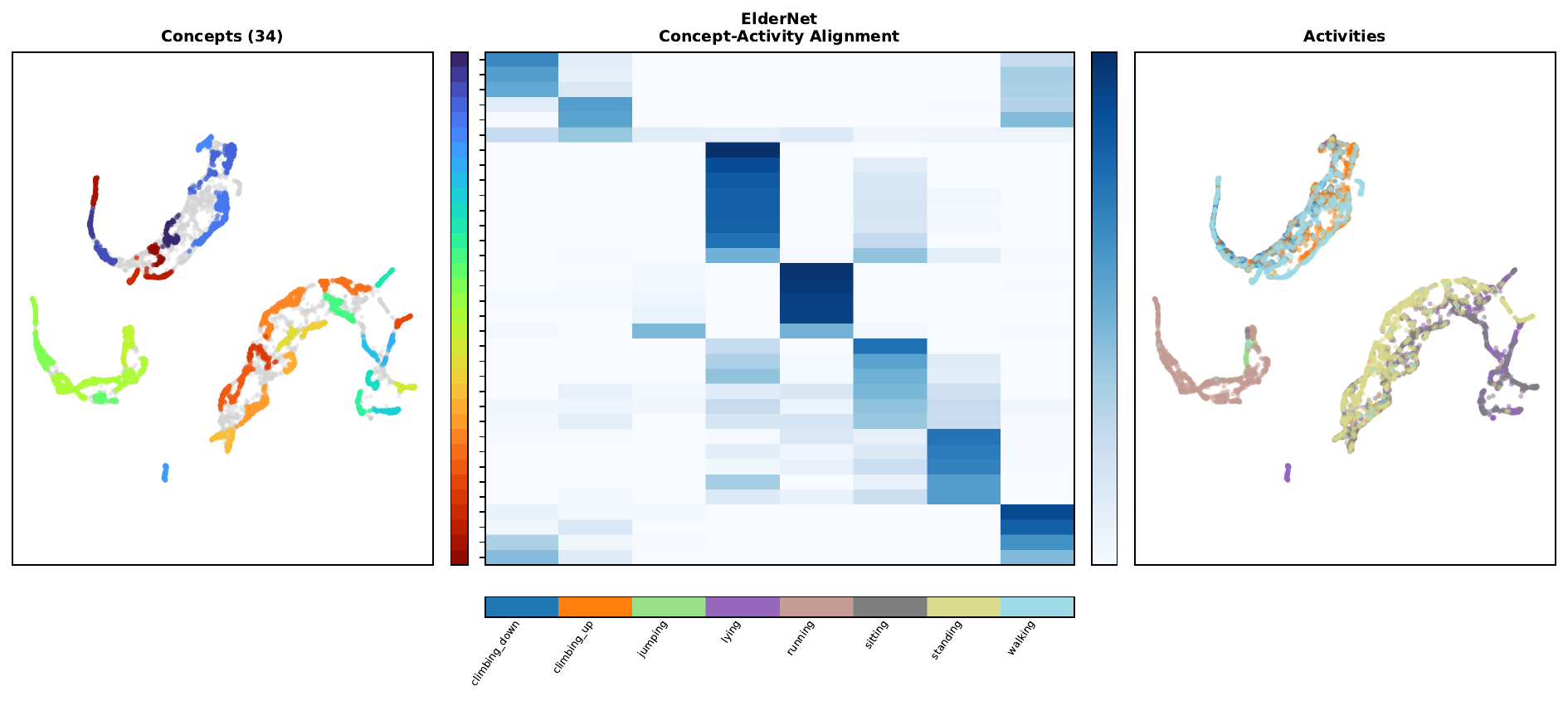}
    \caption{Concept-activity alignment for pretrained ElderNet on RealWorld, based on 107 concepts discovered by NLMCD from the last-layer representations. Left: UMAP of the discovered concepts. Center: concept activation proportions per ground-truth activity label (row-normalized). Right: UMAP of the same samples colored by ground-truth activity.} 
    \label{fig:UMAP_eldernet} 
    %\Description{UMAPs show the same clusters for the concepts and activities just colored differently. Center heatmap shows the distribution of the concepts on ground truth labels, ranging from 0 (no concept activation) to highest (1.0).}
\end{figure*}

Fig.~\ref{fig:UMAP_eldernet} shows the concept-activity alignment for ElderNet, the model with the most comprehensive HAR representations, on the RealWorld dataset. NLMCD discovers 34 concepts from the last layer representations. Dynamic activities such as jumping and running tend to show the strongest activation of a small number of individual concepts with lesser overlap to concepts used for other activities. In contrast walking-related activities such as climbing up or down show diffuse concept activation, overlapping strongly with concepts activated for walking. This suggests that the model does not form clearly separated representations for these activities. Static activities such as lying, sitting or standing tend to have strong overlaps of activated concepts. This pattern is consistent across all investigated FMs (not shown).%Appendix Fig.~\ref{fig:UMAP_normwear}, \ref{fig:UMAP_unimts}, and \ref{fig:UMAP_oxfordssl}).
The public video recordings for RealWorld show that the climbing up and down activities consisted of walking bouts up-/downhill, up-/downstairs , turns, standing and mixed usage of handrails, so ambiguity between those activities is to be expected due to the aspects of data recording under real conditions.

We see these results as a promising sign for concepts discovered from FM feature representations to complement or replace man-made HAR activity labels for a data-driven characterization of activity profiles from large-scale wearables data. We envision a validation of the discovered concepts in comparison to the video recordings from the RealWorld dataset as promising direction for future research.

\subsection{Comparative assessment and recommendations for practitioners}
\label{recommendation}

\begin{table*}[!ht]
    \centering
    \caption{Median statistical rankings across evaluation modes by categories. Rankings (Finetune/Linear/Frozen) represent the median performance position across all datasets within each category. Lower values indicate better overall performance. Overall rank reflects median of the five categories. To guide the eye, we highlight the highest ranked model in each category in bold face. Demo. = demographic datasets.}
    \begin{tabular}{ccccc|cccc}
        \toprule
        \textbf{Dataset} & \textbf{Normwear} & \textbf{UniMTS} & \textbf{Oxford} & \textbf{ElderNet} & \textbf{TinyHAR} & \textbf{Inception} & \textbf{S4} & \textbf{XResNet}\\
        \midrule
        HAR & 2/3/3.5 & 4/\textbf{1}/\textbf{1} & \textbf{1}/3/4 & \textbf{1}/2.5/5 & 3/4/3 & 2/\textbf{1}/2.5 & 2/\textbf{1}/1.5 & 2/2.5/2.5 \\
        Fall & 7.5/7.5/5.5 & 2/\textbf{2.5}/2.5 & \textbf{1}/3/\textbf{1.5} & 3/3/\textbf{1.5} & 6/5.5/6 & 5/4/5 & 2/\textbf{2.5}/2.5 & 3/3/3 \\
        Sleep & 4/8/\textbf{1} & \textbf{1}/\textbf{1}/\textbf{1} & 4/\textbf{1}/4 & 4/\textbf{1}/4 & 4/\textbf{1}/4 & \textbf{1}/\textbf{1}/\textbf{1} & \textbf{1}/\textbf{1}/4 & 4/\textbf{1}/4 \\
        Stress & 3/3/5 & 7/6/5 & \textbf{1}/\textbf{1}/\textbf{1} & \textbf{1}/\textbf{1}/\textbf{1} & 7/8/8 & 3/3/3 & 3/3/3 & 3/6/5 \\
        Demographics & 3/\textbf{2}/\textbf{1} & 4/4/4 & 5/6/4 & 4/6/5 & 7/6/6 & \textbf{2}/3/3 & 5/4/5 & 3/\textbf{2}/3 \\
        \midrule
        \textbf{Median} & 3/4/2 & 3/3/1 & 1/5/4 & 2/4/4 & 5/5/6 & 3/3/3 & 3/3/4 & 3/2/4 \\
        \bottomrule
    \end{tabular}
    \label{tab:combined_eval_results_median_ranks}
\end{table*}

We summarize the achieved results in Table~\ref{tab:combined_eval_results_median_ranks}, which lists the performance ranks aggregated according to task categories. 
The FMs' capabilities are highly heterogeneous across task categories, and the most suitable FM depends primarily on the availability of labeled data and on whether the model is used as a feature extractor or under a computational budget. When labeled data is available, Oxford SSL and ElderNet are the strongest FMs, achieving the best overall ranks (1 and 2) and leading on HAR, falls and stress. In addition, both are also among the most efficient models (Table \ref{tab:model_efficiency}), which makes them the preferred FMs for stress detection. 
This downstream adaptability, however, does not translate to their pretrained representations: as frozen feature extractors both fall behind the supervised baselines. Complementary, UniMTS is the best frozen feature extractor overall, since it is the only FM whose frozen representations rank ahead of the strongest baselines (Inception1D, XResNet1D). Although it ranks last among the FMs on HAR under finetuning, 
it is the most consistent choice for fall detection across evaluation modes. 
NormWear performs on par with the supervised baselines across all tasks and benefits the most from the non-linear head, which makes it best suited to extract demographic properties of the subjects; this comes, however, at a substantial computational cost, roughly two orders of magnitude above the baselines, limiting it to offline processing. 
For sleep staging, no model can be recommended, as absolute performance stays near chance. Contrary to expectations, even FMs that have been pretrained on different sensor positions, fail to beat supervised baselines on specific sensor positions. Overall FMs seem to perform at least similarly good or better (ElderNet) as baseline models (with NormWear the exception) for HAR detection, making them a better choice in scenarios where sensor robustness is required. In summary, supervised baselines such as Inception1D and XResNet1D are sufficient and considerably more efficient if the target task involves well-defined activity classes, fall detection or robustness to sensor placement. In contrast, FMs add value mainly as frozen extractors with a non-linear head, led by UniMTS overall and NormWear for demographics.

\section{Summary and Conclusions}
We benchmarked four FMs and four supervised baselines across HAR, fall and stress detection, sleep staging and demographics (age, weight and height). This benchmark includes efficiency analysis, sensor position adaptability and representational analysis via concept discovery. There is no single dominating model, neither in the supervised nor in the FM domain, but different models perform strongly in different categories or even specific datasets. Even in the most widely considered downstream domain of HAR, FMs do not show consistent advantages over models trained from scratch\, leaving clear room for improvement in FM development. Where FMs add value is task- and evaluation-mode-specific.
Overall, the supervised baselines remain the more efficient and equally or more performant choice for standard HAR, whereas FMs are most useful for fall and stress detection, demographic characterization, robustness to sensor placement (ElderNet), and as ready-to-use feature extractors (UniMTS). 
To foster reproducibility, the source code underlying this study is publicly available at \url{https://github.com/AI4HealthUOL/movement-fm-benchmarking}.

%\appendices

%Appendixes, if needed, appear before the acknowledgment.

\section*{References}

%\subsection{References}
\bibliographystyle{IEEEtran}
\bibliography{sample-base}
\appendix

\subsection{Foundation Models}
We investigate four foundation models, which are described below and also summarized in Table~\ref{tab:models}.

\textbf{UniMTS}
The Unified pre-trained model for Motion Time Series (UniMTS) addresses three challenges that is device placement, device orientation changes during data collection and generalizes activities. The model was pretrained on publicly available HumanML3D \cite{guoGeneratingDiverseNatural2022} dataset that was augmented for multiple sensor positions and augmented text. The model architecture comprises a graph encoder that maps the sensor locations to the signals and a text encoder that generates the activity classes \cite{zhangUniMTSUnifiedPretraining2024}.

\textbf{NormWear}
A normative foundation model that was pre-trained on 14,943 hours of data from nine publicly available multimodal time-series datasets (comprises photoplethysmography, electrocardiography, galvanic skin response and inertial measurement unit (IMU) modalities) and evaluated on 18 publicly available downstream tasks \cite{luoFoundationModelMultivariate2024}. The model tries to integrate multivariate signals with a strong focus on variable channels and sensor types. The main focus of the models data input lies on smart health monitoring. Its architecture consists of a convolutional patching layer and 12 transformer blocks, the decoder consists of two transformer blocks followed by a linear projection and a convolutional layer. Each input channel uses its own [CLS] token.

\textbf{Oxford SSL}
The Oxford SSL model is a self-supervised foundation model for human activity recognition, pre-trained on approximately 700,000 person-days of free-living accelerometer data from over 100,000 participants in the UK Biobank study \cite{yuanSelfsupervisedLearningHuman2024}. Participants wore wrist-worn triaxial accelerometers (Axivity AX3) for seven days during their usual daily activities, real natural human movements outside laboratory settings. The model employs multi-task self-supervised learning with three pretext tasks: arrow of time (detecting signal reversal), permutation (identifying shuffled signal segments), and time warping (recognizing temporally distorted signals). The architecture is based on ResNet-V2 with 18 layers adapted for 1D convolutions, producing a 1024-dimensional feature vector. Data was processed in 10-second windows at 30~Hz sampling rate. Weighted sampling was applied during training.

\textbf{ElderNet}
ElderNet is a gait detection model built upon the UK Oxford SSL foundation model, specifically adapted and optimized for older adults with and without gait impairments. The model addresses the limited availability of labeled daily living data for geriatric populations. ElderNet extends the Oxford SSL model through a two-stage approach: in the first stage, the pre-trained Oxford SSL ResNet-V2 model served as a frozen feature extractor, with additional fully-connected layers added on top. These additional layers were trained using multi-task self-supervised learning on unlabeled data from the Rush Memory and Aging Project (MAP), comprising over 1,000 older adults with and without impaired gait who wore GENEActiv wrist accelerometers for up to 10 days \cite{bennettOverviewFindingsRush2012}. The pretext tasks include predicting signal augmentations (reversal, permutation, time warping, and scaling). In the second stage, the entire model was fine-tuned on labeled data from the Mobilise-D technical validation study \cite{mico-amigoAssessingRealworldGait2023b} comprising 83 participants across five clinical cohorts (Parkinson's disease, proximal femoral fracture, COPD, congestive heart failure, and healthy adults) \cite{brandSelfsupervisedLearningWristworn2024}. In this study, we used the pretrained SSL checkpoint.

\subsection{Supervised Baselines}
In addition to the foundation models discussed so far, we also included supervised three baselines that have shown strong performance in physiological time series classification tasks in addition to a modality-specific baseline.

\textbf{TinyHAR}
TinyHAR is a state of the art lightweight designed specifically for sensor-based HAR. Its architecture consist of convolutional layers for feature extraction, a transformer encoder for cross-channel interaction, a fully connected for feature fusion, a single LSTM layer for temporal information extraction and a self attention module to enhance temporal information \cite{zhouTinyHARLightweightDeep2022}.

\textbf{XResNet1D} The XResNet1D50 model is a one-dimensional adaptation of ResNet-based convolutional neural network (CNN), proposed originally in the context of ECG classification \cite{strodthoff2020deep}, which has been successfully applied to other application domains such as classification and regression task on PPG data \cite{moulaeifard2025machine}.

\textbf{Inception1D} The Inception1D represents another widely used CNN model, which was identified as particularly well-adapted model for time series classification tasks \cite{ismail2020inceptiontime}.

\textbf{S4} The structured state space sequence (S4) model represents a qualitatively different supervised baseline model, which leverages structured state space models \cite{guefficiently}. It consistently outperformed CNN-based baselines in ECG classification \cite{al2025benchmarking} and yielded best-performing results for EEG-based sleep staging \cite{wang2025s4sleep}.

\subsection{Datasets and tasks}
For a comprehensive evaluation, we leverage a large number of publicly available dataset, which are described below. We summarize the datasets in Table~\ref{tab:datasets}.

\textbf{PAMAP2}
The Physical Activity Monitoring for Aging People (PAMAP2) dataset consists of multimodal inertial and physiological sensor data collected from wrist, chest, and ankle-mounted IMUs at 100~Hz, together with heart rate signals (9~Hz), while 9 subjects performed 18 physical activities under a controlled protocol \cite{reissIntroducingNewBenchmarked2012}. Since the 9th participant has mostly other/transient activities that was not more specified, this participant was for our comparison removed.

\textbf{HAR70+}
Human Activity Recognition 70+ (HAR70+) dataset contains accelerometer and gyroscop data recorded with Axivity AX3 from the lower back and thigh with a sampling frequency of 50~Hz and a range of $\pm8$ g. For labeling the activities, video was recorded as well and synchronized with the other sensors. The subjects followed a semi-structured free-living protocol \cite{ustadValidationActivityType2023}.

\textbf{HARTH}
The human activity recognition Trondheim (HARTH) dataset was recorded in a semi-structured free-living setting using two tri-axial Axivity AX3 accelerometers with 100~Hz and later downsampled to 50~Hz with a range of $\pm8$ g (sensor location was lower back and upper thigh). In total 37~h (22 participants) were recorded and labeled using data from video camera that pointed downwards to record leg movements \cite{logacjovHARTHHumanActivity2021}.

\textbf{RealWorld2016}
The RealWorld2016 dataset contains multimodal sensor data recorded from 15 participants performing 8 activities. Data were collected simultaneously from 7 body positions (chest, forearm, head, shin, thigh, upper arm, and waist) using smartphones and a smartwatch equipped with triaxial accelerometers sampled at 50~Hz. Each activity was recorded for approximately 10 minutes per participant, except jumping ($\sim$1.7 minutes) due to physical exertion. Synchronized video recordings facilitated annotation. The dataset also includes GPS, gyroscope, light, magnetic field, and sound level data \cite{sztylerOnbodyLocalizationWearable2016}.

\textbf{USC-HAD}
The University of Southern California Human Activity Dataset (USC-HAD) contains data from 14 participants performing 12 activities under a supervised protocol. Data was recorded using a triaxial accelerometer ($\pm6$ g) and triaxial gyroscope ($\pm$500$^\circ$/s) at 100~Hz, worn at the front right hip. Each participant performed 5 trials per activity on different days \cite{zhangUSCHADDailyActivity2012a}.

\textbf{WISDM}
The wireless sensor data mining (WISDM) dataset contains accelerometer and gyroscope data collected simultaneously from a smartphone (pocket) and smartwatch (dominant hand) at 20 Hz while 51 subjects performed 18 activities of daily living \cite{kwapiszActivityRecognitionUsing2011a} and later a newer version was updated \cite{weissWISDMSmartphoneSmartwatch}. In the dataset activities are categorized into non-hand-oriented locomotion, general-hand-orientated activities, and eating-related activities. For this benchmark, we focused on non-hand-oriented locomotions. Each activity was performed for three minutes per subject under a controlled protocol.

\textbf{SisFall}
The SisFall dataset contains wearable inertial sensor data for fall detection, collected from 38 participants, including 23 young adults (19--30 years) and 15 elderly adults (60--75 years). Data were recorded using a waist-mounted wearable device equipped with a triaxial accelerometer and a triaxial gyroscope. Accelerometer signals were sampled at 200~Hz with a measurement range of $\pm16$ g. The dataset comprises 19 activities of daily living (ADLs) and 15 types of simulated falls, including forward, backward, and lateral falls from standing, walking, and sitting. Elderly participants performed ADLs only, with no fall trials recorded, except for a single elderly subject (SE06) who simulated both ADLs and all fall types; all remaining fall recordings were performed by young adults \cite{sucerquiaSisFallFallMovement2017a}.

\textbf{KFall}
The KFall dataset contains a large variety of activities of daily living and 15 types of simulated falls. Data were recorded from 32 healthy young adult participants using a nine-axis inertial sensor (LPMS-B2), comprising a triaxial accelerometer ($\pm16$ g), a triaxial gyroscope ($\pm2000^\circ$/s), and a triaxial magnetometer ($\pm$16 g). All inertial signals were sampled at 100~Hz. In addition, synchronized video recordings at 90~fps were acquired and used for frame-level annotation of fall onset and fall impact times. The inertial sensor was worn on the lower back, resulting in time-aligned multimodal motion recordings \cite{yuLargescaleOpenMotion2021}.

\textbf{Dualsleep}
The DualSleep dataset comprises triaxial accelerometry and skin temperature data recorded from lower back and thigh-mounted Axivity AX3 sensors in 29 adults. Accelerometer data were sampled at 100~Hz and downsampled to 50~Hz, while skin temperature was recorded at 1.2~Hz. Sleep and wakefulness were annotated using synchronized polysomnography scored in 30~s epochs, yielding time-aligned multimodal recordings \cite{logacjovMachineLearningModel2024}.

\textbf{WESAD}
The Wearable Stress and affect detection dataset (WESAD). It contains multimodal data from a wrist worn Empatica E4 and a chest worn RespiBAN Professional. The later recorded with 700 Hz ECG, EDA, EMG and TEMP, whereas the wrist worn device recorded BVP (64 Hz), EDA (4 Hz), TEMP (4 Hz) and triaxial ACC (32 Hz) \cite{schmidtIntroducingWESADMultimodal2018}.

\subsection{Additional results}

\begin{figure*}[b]
    \centering
    \includegraphics[width=1\linewidth]{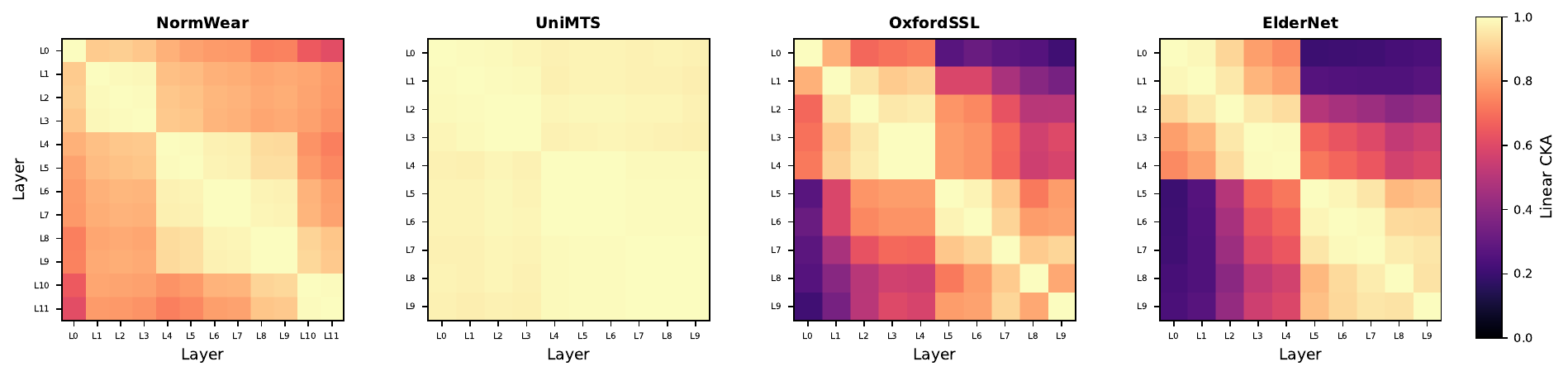}
    \caption{Intra-model layerwise CKA of Foundation models on the RealWorld dataset using the pretrained checkpoints. Lighter colors indicate higher linear similarity between representations. Implementation of linear CKA followed \cite{kornblith2019similarity}. This is the pretrained counterpart to the finetuned CKA shown in Fig.~\ref{fig:cka_fms_finetuned}}.
    \label{fig:cka_fms}
    %\Description{Four heatmaps showing pairwise linear CKA similarity across layers for each of the following models: NormWear, UniMTS, Oxford SSL, and ElderNet.}
\end{figure*}

\begin{figure*}
    \centering
    \includegraphics[width=1\linewidth]{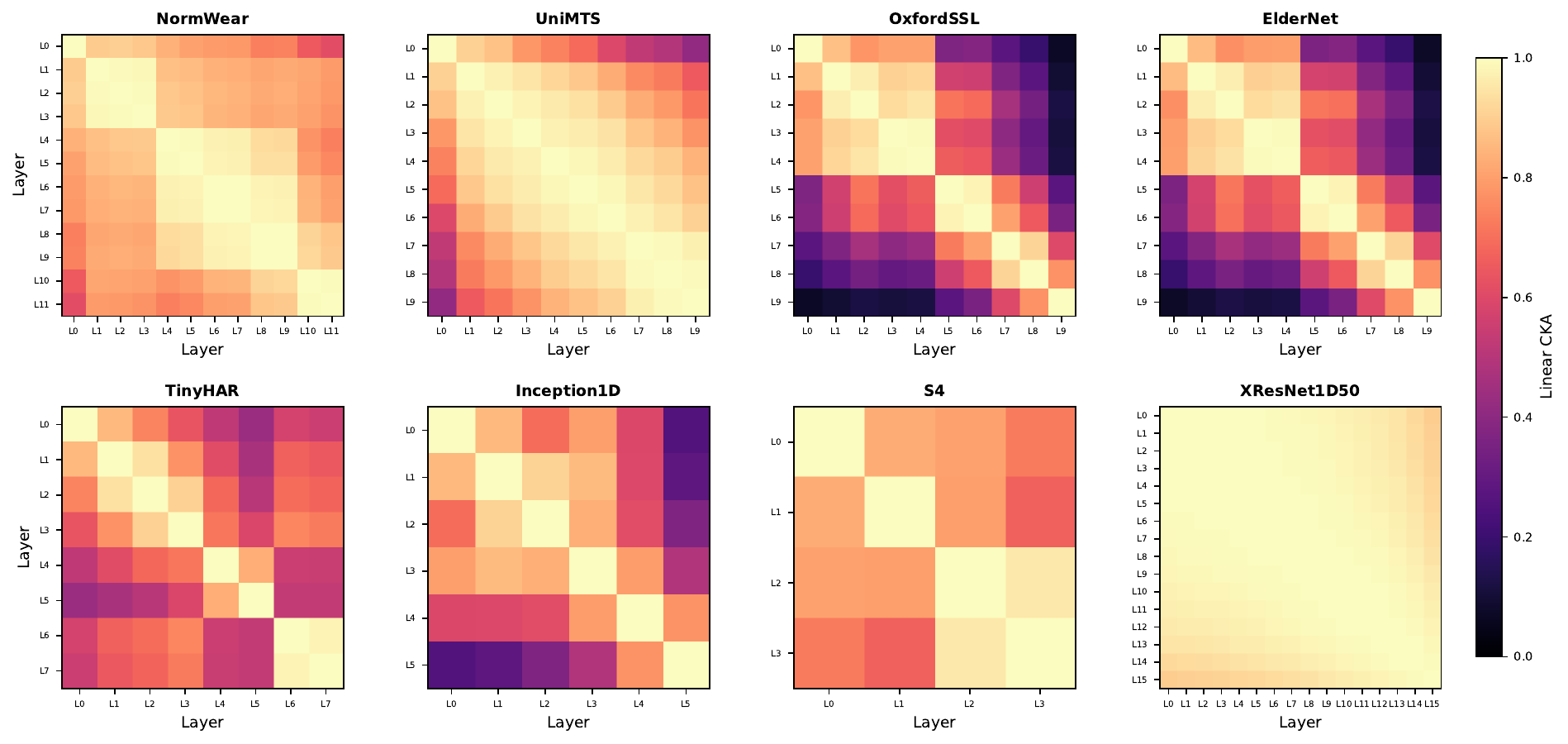}
    \caption{Intra-model CKA on RealWorld (wrist) for the four FMs and baseline models after finetuning. Lighter colors indicate higher linear similarity between the representations of layer pairs within the same model. Implementation of linear CKA followed \cite{kornblith2019similarity}.}
    \label{fig:cka_fms_finetuned}
    %\Description{UMAPs show the same clusters for the concepts and activities just colored differently. Center heatmap shows the distribution of the concepts on ground truth labels, ranging from 0 (no concept activation) to highest (1.0).}
\end{figure*}

Tables \ref{tab:full_finetune} -- \ref{tab:non_lin_frozen} show the evaluation results across the different modes for the FMs and baseline models. %Table \ref{tab:combined_eval_results} shows the the overall rank across all evaluation modes.
Fig. \ref{fig:cka_fms} and \ref{fig:cka_fms_finetuned} show the layerwise intra-model CKA for the FMs before finetuning and for the FMs and baselines after finetuning on RealWorld wrist dataset, respectively.
UniMTS maintains near-uniform CKA across layers, NormWear shows a clear representational shift in its earliest and deepest layers, while Oxford SSL and ElderNet show similar block structures between early (up to L4) and later layers. The similarity between early and later layers is more gradually changing across layers for Oxford SSL compared to ElderNet, whereas the early or later layers in ElderNet are more homogeneous within each other.
Additional concept-activity alignments for the remaining FMs can be found online \url{https://github.com/AI4HealthUOL/movement-fm-benchmarking/tree/main/concept_analysis/results/paper_figures}.
%Fig. \ref{fig:UMAP_normwear}--\ref{fig:UMAP_oxfordssl} show the concept-activity alignments for NormWear, UniMTS and Oxford SSL, respectively.

\begin{table*}
    \centering
    \caption{Finetuning evaluation results of the four FMs (left block) and four supervised baselines (right block) per dataset. Classification tasks (HAR, Sleep, Fall, Stress) are reported as AUROC (higher is better), whereas demographic tasks (Age, Weight, Height) are reported as MAE (lower is better). Per dataset, the best overall model is bold and underlined, models who rank statistical tied to the best model are bold.}%Finetune evaluation results}
    \begin{tabular}{ccccc|cccc}
    \toprule
        \textbf{Dataset} & \textbf{NormWear} & \textbf{UniMTS} & \textbf{Oxford} & \textbf{ElderNet} & \textbf{TinyHAR} & \textbf{Inception} & \textbf{S4} & \textbf{XResNet}\\
        \midrule
        \textbf{HAR}\\
        PAMAP2 & 0.874 & 0.886 & \textbf{0.887} & 0.884 & \textbf{0.917} & \textbf{0.913} & 0.874 & \textbf{\underline{0.918}} \\
        HAR70+ & \textbf{\underline{0.943}} & 0.915 & 0.840 & 0.879 & 0.907 & 0.888 & \textbf{0.932} & 0.849 \\
        HARTH & \textbf{0.981} & \textbf{0.987} & \textbf{0.984} & \textbf{0.984} & 0.963 & \textbf{0.983} & \textbf{\underline{0.987}} & \textbf{0.987} \\
        USC-HAD & \textbf{0.953} & \textbf{0.957} & 0.939 & \textbf{0.957} & \textbf{0.958} & \textbf{0.966} & \textbf{\underline{0.970}} & \textbf{0.964} \\
        WISDM & 0.917 & 0.938 & \textbf{0.955} & \textbf{\underline{0.957}} & 0.945 & 0.941 & 0.947 & 0.926 \\
        RealWorld & 0.937 & 0.865 & \textbf{0.958} & \textbf{\underline{0.959}} & 0.894 & 0.887 & 0.852 & 0.896 \\
        \midrule
        \textbf{Sleep}\\
        Dualsleep & 0.589 & \textbf{0.619} & 0.564 & 0.524 & 0.590 & \textbf{\underline{0.621}} & \textbf{0.620} & 0.591 \\
        \midrule
        \textbf{Fall}\\
        KFall & 0.956 & \textbf{\underline{0.990}} & \textbf{0.985} & 0.984 & 0.948 & 0.983 & \textbf{0.988} & \textbf{0.986} \\
        %KFall binary & \textbf{0.996} & \textbf{1.000} & \textbf{0.999} & \textbf{0.999} & \textbf{0.994} & \textbf{\underline{1.000}} & \textbf{1.000} & \textbf{0.992} \\
        SisFall & 0.889 & 0.914 & \textbf{0.983} & \textbf{\underline{0.985}} & 0.900 & 0.910 & 0.921 & 0.908 \\
        %SisFall binary & 0.965 & 0.971 & \textbf{0.999} & \textbf{\underline{0.999}} & 0.984 & 0.974 & 0.974 & 0.969 \\
        \midrule
        \textbf{Stress}\\
        WESAD & 0.568 & 0.477 & \textbf{0.650} & \textbf{\underline{0.665}} & 0.457 & 0.525 & 0.567 & 0.509 \\
        \midrule
        \textbf{Demographics}\\
        \midrule
        \textbf{Age (years)}\\
        PAMAP2 & 7.116 & 7.710 & 7.296 & 6.509 & 7.006 & \textbf{\underline{6.409}} & 6.756 & 9.173 \\
        SisFall & 14.096 & \textbf{10.149} & 12.277 & 13.319 & 15.763 & \textbf{10.289} & \textbf{\underline{8.770}} & \textbf{9.805} \\
        USC-HAD & 12.874 & 14.431 & \textbf{\underline{7.718}} & 14.293 & 12.927 & 14.763 & 14.380 & 17.451 \\
        WESAD & 1.217 & \textbf{\underline{0.260}} & 0.909 & 0.539 & 1.519 & 0.667 & 1.028 & 0.394 \\
        \midrule
        \textbf{Weight (kg)}\\
        PAMAP2 & 11.189 & 11.666 & \textbf{\underline{9.504}} & 12.399 & 47.470 & 13.541 & 9.667 & 11.243 \\
        USC-HAD & 28.055 & 30.319 & \textbf{\underline{5.155}} & 16.919 & 28.424 & 30.754 & 29.612 & 27.196 \\
        WESAD & 19.484 & 15.599 & \textbf{\underline{3.518}} & 16.141 & 18.505 & 15.062 & 18.326 & 15.596 \\
        \midrule
        \textbf{Height (cm)}\\
        USC-HAD & 13.289 & 12.303 & 14.320 & \textbf{\underline{11.050}} & 118.542 & 16.325 & 14.215 & 13.808 \\
        WESAD & \textbf{3.921} & \textbf{3.943} & 6.514 & 10.190 & 56.343 & \textbf{\underline{3.909}} & 10.387 & \textbf{3.932} \\
    \bottomrule
    \end{tabular}
    \label{tab:full_finetune}
\end{table*}

\begin{table*}
    \centering
    \caption{Linear evaluation results of the four FMs (left block) and four supervised baselines (right block) per dataset. Classification tasks (HAR, Sleep, Fall, Stress) are reported as AUROC (higher is better), whereas demographic tasks (Age, Weight, Height) are reported as MAE (lower is better). Per dataset, the best overall model is bold and underlined, models who rank statistical tied to the best model are bold.}
    \begin{tabular}{ccccc|cccc}
    \toprule
        \textbf{Dataset} & \textbf{NormWear} & \textbf{UniMTS} & \textbf{Oxford} & \textbf{ElderNet} & \textbf{TinyHAR} & \textbf{Inception} & \textbf{S4} & \textbf{XResNet}\\
        \midrule
        \textbf{HAR}\\
        PAMAP2 & 0.874 & \textbf{0.905} & \textbf{0.895} & \textbf{0.894} & \textbf{0.917} & \textbf{0.913} & 0.874 & \textbf{\underline{0.918}} \\
        HAR70+ & \textbf{\underline{0.943}} & \textbf{0.925} & 0.841 & 0.855 & 0.907 & 0.888 & \textbf{0.932} & 0.849 \\
        HARTH & \textbf{0.980} & 0.979 & 0.968 & 0.971 & 0.963 & \textbf{0.983} & \textbf{\underline{0.987}} & \textbf{0.987} \\
        USC-HAD & 0.953 & \textbf{\underline{0.972}} & 0.954 & 0.953 & 0.958 & \textbf{0.966} & \textbf{0.970} & \textbf{0.964} \\
        WISDM & 0.917 & \textbf{0.945} & \textbf{0.950} & \textbf{\underline{0.950}} & \textbf{0.945} & \textbf{0.941} & \textbf{0.947} & 0.926 \\
        RealWorld & \textbf{0.935} & 0.834 & \textbf{0.949} & \textbf{\underline{0.950}} & 0.894 & 0.887 & 0.852 & 0.896 \\
        \midrule
        \textbf{Sleep}\\
        Dualsleep & 0.569 & \textbf{\underline{0.621}} & \textbf{0.600} & \textbf{0.569} & \textbf{0.590} & \textbf{0.621} & \textbf{0.620} & \textbf{0.591} \\
        \midrule
        \textbf{Fall}\\
        KFall & 0.957 & \textbf{\underline{0.993}} & 0.971 & 0.971 & 0.948 & 0.983 & 0.988 & 0.986 \\
        SisFall & 0.890 & 0.912 & \textbf{0.980} & \textbf{\underline{0.981}} & 0.900 & 0.910 & 0.921 & 0.908 \\
        \midrule
        \textbf{Stress}\\
        WESAD & 0.577 & 0.479 & \textbf{0.665} & \textbf{\underline{0.665}} & 0.457 & 0.525 & 0.567 & 0.509 \\
        \midrule
        \textbf{Demographics}\\
        \midrule
        \textbf{Age (years)}\\
        PAMAP2 & 7.224 & 8.734 & 9.931 & 9.536 & 7.006 & \textbf{\underline{6.409}} & 6.756 & 9.173 \\
        SisFall & \textbf{13.873} & \textbf{9.873} & 15.720 & 15.625 & 15.763 & \textbf{10.289} & \textbf{\underline{8.770}} & \textbf{9.805} \\
        USC-HAD & 13.187 & 14.073 & 15.259 & 15.611 & \textbf{\underline{12.927}} & 14.763 & 14.380 & 17.451 \\
        WESAD & 1.322 & \textbf{\underline{0.248}} & 1.006 & 0.724 & 1.519 & 0.667 & 1.028 & \textbf{0.394} \\
        \midrule
        \textbf{Weight (kg)}\\
        PAMAP2 & 11.168 & 13.162 & 35.212 & 34.665 & 47.470 & 13.541 & \textbf{\underline{9.667}} & 11.243 \\
        USC-HAD & 28.135 & 28.151 & 23.960 & \textbf{\underline{22.554}} & 28.424 & 30.754 & 29.612 & 27.196 \\
        WESAD & 19.090 & 17.795 & 15.083 & \textbf{\underline{13.762}} & 18.505 & 15.062 & 18.326 & 15.596 \\
        \midrule
        \textbf{Height (cm)}\\
        USC-HAD & 13.143 & \textbf{\underline{12.912}} & 97.160 & 98.607 & 118.542 & 16.325 & 14.215 & 13.808 \\
        WESAD & \textbf{\underline{3.902}} & 4.098 & 51.977 & 51.381 & 56.343 & \textbf{3.909} & 10.387 & \textbf{3.932} \\
    \bottomrule
    \end{tabular}
    \label{tab:lin_frozen}
\end{table*}

\begin{table*}
    \centering
    \caption{Frozen evaluation results of the four FMs (left block) and four supervised baselines (right block) per dataset. Classification tasks (HAR, Sleep, Fall, Stress) are reported as AUROC (higher is better), whereas demographic tasks (Age, Weight, Height) are reported as MAE (lower is better). Per dataset, the best overall model is bold and underlined, models who rank statistical tied to the best model are bold.}
    \begin{tabular}{ccccc|cccc}
    \toprule
        \textbf{Dataset} & \textbf{NormWear} & \textbf{UniMTS} & \textbf{Oxford} & \textbf{ElderNet} & \textbf{TinyHAR} & \textbf{Inception} & \textbf{S4} & \textbf{XResNet}\\
        \midrule
        \textbf{HAR}\\
        PAMAP2 & 0.886 & \textbf{0.899} & 0.884 & 0.885 & \textbf{0.917} & \textbf{0.913} & 0.874 & \textbf{\underline{0.918}} \\
        HAR70+ & 0.930 & \textbf{\underline{0.961}} & 0.874 & 0.881 & 0.907 & 0.888 & 0.932 & 0.849 \\
        HARTH & 0.971 & \textbf{0.985} & \textbf{0.978} & 0.976 & 0.963 & \textbf{0.983} & \textbf{\underline{0.987}} & \textbf{0.987} \\
        USC-HAD & \textbf{0.963} & \textbf{\underline{0.972}} & 0.953 & 0.947 & 0.958 & \textbf{0.966} & \textbf{0.970} & \textbf{0.964} \\
        WISDM & 0.931 & \textbf{0.951} & \textbf{\underline{0.951}} & \textbf{0.950} & \textbf{0.945} & 0.941 & \textbf{0.947} & 0.926 \\
        RealWorld & \textbf{\underline{0.967}} & 0.889 & 0.953 & \textbf{0.955} & 0.894 & 0.887 & 0.852 & 0.896 \\
        \midrule
        \textbf{Sleep}\\
        Dualsleep & \textbf{0.623} & \textbf{\underline{0.638}} & 0.555 & 0.546 & 0.590 & \textbf{0.621} & 0.620 & 0.591 \\
        \midrule
        \textbf{Fall}\\
        KFall & 0.967 & \textbf{\underline{0.994}} & 0.987 & 0.987 & 0.948 & 0.983 & 0.988 & 0.986 \\
        SisFall & 0.904 & 0.904 & \textbf{0.985} & \textbf{\underline{0.986}} & 0.900 & 0.910 & 0.921 & 0.908 \\
        \midrule
        \textbf{Stress}\\
        WESAD & 0.493 & 0.469 & \textbf{\underline{0.657}} & \textbf{0.643} & 0.457 & 0.525 & 0.567 & 0.509 \\
        \midrule
        \textbf{Demographics}\\
        \midrule
        \textbf{Age (years)}\\
        PAMAP2 & \textbf{\underline{6.274}} & 7.476 & 6.641 & 6.859 & 7.006 & 6.409 & 6.756 & 9.173 \\
        SisFall & 14.457 & \textbf{11.094} & 12.685 & 13.270 & 15.763 & \textbf{10.289} & \textbf{\underline{8.770}} & \textbf{9.805} \\
        USC-HAD & 13.904 & \textbf{\underline{4.995}} & 13.071 & 12.773 & 12.927 & 14.763 & 14.380 & 17.451 \\
        WESAD & \textbf{0.472} & 0.819 & 1.093 & 0.795 & 1.519 & \textbf{0.667} & 1.028 & \textbf{\underline{0.394}} \\
        \midrule
        \textbf{Weight (kg)}\\
        PAMAP2 & 9.949 & 12.449 & 11.306 & 11.534 & 47.470 & 13.541 & \textbf{\underline{9.667}} & 11.243 \\
        USC-HAD & \textbf{\underline{10.642}} & \textbf{12.799} & 17.528 & 17.584 & 28.424 & 30.754 & 29.612 & 27.196 \\
        WESAD & \textbf{\underline{2.388}} & 3.144 & 18.964 & 19.782 & 18.505 & 15.062 & 18.326 & 15.596 \\
        \midrule
        \textbf{Height (cm)}\\
        USC-HAD & \textbf{\underline{8.690}} & 12.625 & 9.780 & 11.512 & 118.542 & 16.325 & 14.215 & 13.808 \\
        WESAD & \textbf{4.273} & 4.989 & 7.188 & 6.946 & 56.343 & \textbf{\underline{3.909}} & 10.387 & \textbf{3.932} \\
    \bottomrule
    \end{tabular}
    \label{tab:non_lin_frozen}
\end{table*}

\end{document}